\documentclass[12pt,oneside]{article}

\usepackage{amsmath}
\usepackage{amscd}
\usepackage{amsfonts}
\usepackage{amsthm}
\usepackage{a4}

\def\IN{\mathbb{N}}
\def\IZ{\mathbb{Z}}
\def\IR{\mathbb{R}}

\newcommand{\tr}{\mbox{${\rm tr \;}$}}

\newcommand{\tail}{\mbox{$({\rm tail})$}}

\begin{document}
\thispagestyle{empty}
\setcounter{page}{0}
\renewcommand{\theequation}{\thesection.\arabic{equation}}

{\hfill{VUB/TENA/01/02}} 

{\hfill{hep-th/0103015}}

{\hfill{March 2, 2001}}

\vspace{2cm}

\begin{center}
{\bf THE UNIQUENESS OF THE ABELIAN BORN-INFELD ACTION}

\vspace{1.4cm}

LIES DE FOSSE, PAUL KOERBER\footnote{Aspirant FWO} and ALEXANDER SEVRIN

\vspace{.2cm}

{\em Theoretische Natuurkunde, Vrije Universiteit Brussel} \\
{\em Pleinlaan 2, B-1050 Brussels, Belgium} \\
\end{center}

\vspace{-.1cm}

\centerline{{\tt lies, koerber, asevrin@tena4.vub.ac.be}}
\vspace{1cm}
\centerline{ABSTRACT}

\vspace{- 4 mm}  

\begin{quote}\small
Starting from BPS solutions to Yang-Mills which define a stable holomorphic 
vector bundle, we investigate its deformations. Assuming slowly varying 
fieldstrengths, we find in the abelian case a {\em unique} deformation given by 
the abelian Born-Infeld action. We obtain the deformed 
Donaldson-Uhlenbeck-Yau stability condition to all orders in $\alpha '$. This 
result provides strong evidence supporting the claim that the only 
supersymmetric deformation of the abelian $d=10$ supersymmetric Yang-Mills 
action is the Born-Infeld action. 
\end{quote}
\baselineskip18pt
\noindent

\vspace{5mm}

\newpage

\setcounter{equation}{0}
\section{Introduction}
An exciting consequence of the discovery of D-branes \cite{pol} was their
close relation to gauge theories. The worldvolume
degrees of freedom of a single D$p$-brane are $9-p$ scalar fields and a
$U(1)$ gauge field in $p+1$ dimensions. The former describe the
transversal fluctuations of the D-brane while the latter describes an
open string longitudinal to the brane. For slowly varying fields, the
effective action governing the low-energy dynamics of a D-brane is known
through all orders in $\alpha '$: it is the ten-dimensional
supersymmetric Born-Infeld action, dimensionally reduced to $p+1$
dimensions \cite{BI}. Its supersymmetric extension was obtained in
\cite{susynbi}. The knowledge of the full effective action was crucial
for numerous applications.

Once several, say $n$, D-branes coincide, the gauge group is enhanced
from $U(1)$ to $U(n)$, \cite{witten}. The non-abelian extension of the
Born-Infeld theory is not known yet. The most natural form for it is the
symmetrized trace proposal in \cite{Tstr}. However, as shown in \cite{HT}
and \cite{DST}, this does not correctly capture all of the D-brane
dynamics. Using the mass spectrum as a guideline, partial higher order
results were obtained in \cite{STT}. In \cite{kappa}, $\kappa$-symmetry
was shown to be a powerful but technically involved tool to fix the
ordenings ambiguities.

In \cite{DST}, it was pointed out that BPS configurations of D$p$-branes
at angles, \cite{Angles1}, \cite{Angles2}, \cite{townsend},
might provide an important
tool to probe the structure of the effective action. Upon T-dualizing we
end up with D$2p$-branes in the presence of constant magnetic background
fields. In the large volume limit ($\alpha '\rightarrow 0)$ the BPS
conditions define a stable holomorphic vector bundle \cite{GSW}.
Moving away from the large volume limit, these conditions receive $\alpha
'$ corrections. As a BPS configuration necessarily solves the equations
of motion, we obtain relations between different orders in $\alpha '$ in
the effective action.

In the present paper we start the exploration of the consequences of this
idea. As we consider the present paper as a ``feasibility study'' we will
make two simplifying assumptions: we work in the limit of slowly varying
fieldstrengths and restrict our attention to the abelian case. The first
assumption is translated by the fact that we will ignore terms containing
derivatives of the fieldstrength. The second assumption is implemented by
taking the magnetic background fields to live in the Cartan subalgebra of
$u(n)$. As a starting point we take the theory in the $\alpha
'\rightarrow 0$ limit. I.e. we take the Yang-Mills action reduced to the
torus of $U(n)$ in the presence of magnetic background fields which
define a stable holomorphic vector bundle. Subsequently we add arbitrary
powers of the fieldstrength to it and demand that the BPS configurations
solve the equations of motion. This problem turns out to have a {\em
unique} solution. The resulting action is precisely the abelian
Born-Infeld action and the stability condition, also known as the
Donaldson-Uhlenbeck-Yau condition \cite{DUY}, acquires $\alpha '$
corrections which are unique as well.

This result provides a serious incentive to extend the analysis to the
much harder non-abelian case \cite{wij}. In addition, there is a
suspicion that, because of the severely restricted form of the
supersymmetry algebra in ten dimensions, the BI action is the only
supersymmetric deformation of abelian Yang-Mills\footnote{This was
suggested by Savdeep Sethi.}. E.g. supersymmetry fixes in the abelian
case uniquely the fourth order term in the BI action \cite{BMT}. As BPS
configurations are intimately related to supersymmetry, we believe that our
present paper lends strong support to this claim.

Our paper is organized as follows. In the next section we review BPS
configurations of D$p$-branes at angles. The subsequent section relates
this to supersymmetric Yang-Mills theory. Using an example, we outline
our strategy in the third section. Section 4 provides the proof of our
assertion. We discuss our results and diverse applications in the final
section. Conventions are given in the first appendix while the second
gathers some useful results concerning the abelian Born-Infeld action.

\setcounter{equation}{0}
\section{BPS configurations from string theory}
Simple BPS configurations of D-branes arise as follows \cite{Angles1}, 
\cite{Angles2}, \cite{townsend}. 
One starts with two coinciding D$p$-branes. Keeping one of them fixed, 
one performs a Lorentz transformation on the other one.
For all boosts and generic rotations, 
all supersymmetry gets broken in this way. 
However there are particular rotations for 
which some of the supersymmetry is preserved. 

Consider two D$p$-branes in the $(1,3,\cdots, 2p-1)$ directions. 
Keeping one of them fixed, rotate the other one subsequently
over an angle $\phi_1$ in the (1\,2) plane, over an angle $\phi_2$ in the 
(3\,4) plane, ..., over an angle $\phi_p$ in the $(2p-1\,2p)$ plane. The 
following table summarizes for various values of $p$ the BPS conditions 
on the angles (taken to be non-zero unless stated otherwise)
and the number of remaining supersymmetries.

\begin{center}
\begin{tabular}{|c|l|c|}\hline\hline
$p$ &BPS condition & susy's\\ \hline\hline
2 &$\phi_1+\phi_2=2\pi n$ & 8\\ \hline
3 &$\phi_1+\phi_2+\phi_3=2\pi n$ & 4\\ \hline
4 &$\phi_1+\phi_2+\phi_3+\phi_4=2\pi n$ & 2\\ \cline{2-3}
  &$\phi_1+\phi_2=2\pi n$, $\phi_3+\phi_4=2\pi m$&4 \\ \cline{2-3}
  &$\phi_1=\phi_2=\phi_3=\phi_4$&6 
\\ \hline\hline
\end{tabular}
\end{center}

\noindent In the table, we took $n,\,m\in\IZ$.

In order to make contact with the Born-Infeld theory, we T-dualize the 
system in the $2, 4, ..., 2p$ directions. In this way, we end up with 
two coinciding D$2p$-branes with magnetic fields turned on. Indeed, having
two D$2p$-branes extended in the 1, 2, ..., 2p directions with magnetic
flux $F_{2i-1\, 2i}$, $i\in\{1,\cdots,p\}$,
\begin{eqnarray}
F_{2i-1\,2i}=\left(  
\begin{array}{cc}
g_i+f_i&0\\
0&g_i-f_i 
\end{array}
\right),
\end{eqnarray}
we can choose a gauge such that the potentials have the form,
\begin{eqnarray}
A_{2i-1}=0,\quad A_{2i}=F_{2i-1\,2i}x^{2i-1}.\label{pot}
\end{eqnarray}
T-dualizing back, we end up with two D$p$-branes
with transversal coordinates given by 
\begin{eqnarray}
X^{2i}=2\pi\alpha 'A_{2i}.\label{trc}
\end{eqnarray}
Using eq. (\ref{pot}) in eq. (\ref{trc}), we recognize the original 
configuration with the two D$p$-branes at angles with the angles given by
\begin{eqnarray}
\phi_i&=&\arctan(2\pi\alpha '(g_i+f_i))-\arctan(2\pi\alpha '(g_i-f_i))\nonumber\\
&=&\arctan\frac{4\pi\alpha 'f_i}
{1+(2\pi\alpha ')^2(g_i^2-f_i^2)}.
\end{eqnarray}
In the table below we translate the BPS conditions on the angles in 
BPS conditions on the fieldstrengths, choosing for simplicity the $U(1)$ 
part to be zero, $g_i=0$.

\begin{center}
\begin{tabular}{|c|l|l|}\hline\hline
$p$ &BPS condition &fieldstrengths\\ \hline\hline
2 &$\phi_1+\phi_2=2\pi n$ & $f_1+f_2=0$\\ \hline
3 &$\phi_1+\phi_2+\phi_3=2\pi n$ & $f_1+f_2+f_3=(2\pi\alpha ')^2f_1f_2f_3$\\ \hline
4 &$\phi_1+\phi_2+\phi_3+\phi_4=2\pi n$ & $f_1+f_2+f_3+f_4=(2\pi\alpha ')^2(f_1f_2f_3+$\\
&&$\qquad f_1f_3f_4+f_1f_2f_4+f_2f_3f_4)$
\\ \cline{2-3}
  &$\phi_1+\phi_2=2\pi n$, $\phi_3+\phi_4=2\pi m$&$f_1+f_2=f_3+f_4=0$\\ \cline{2-3}
  &$\phi_1=\phi_2=\phi_3=\phi_4$&$f_1=f_2=f_3=f_4$ 
\\ \hline\hline
\end{tabular}
\end{center}

\noindent One notices that for $p>2$, the BPS condition expressed
in terms of fieldstrengths corresponding to the angular relation $\sum_i\,\phi_i=2\pi n$, 
gets $\alpha '{}^2$ corrections. In the next, except 
when stated otherwise, we will always study BPS conditions of this type. 
In the remainder of this paper, we will put $2\pi\alpha '=1$.

\setcounter{equation}{0}
\section{BPS configurations in supersymmetric Yang-Mills}
The supersymmetric $U(n)$ Yang-Mills theory in $d=10$ is given 
by\footnote{We ignore an overall multiplicative constant.}
\begin{eqnarray}
{\cal S}=\int d^{10}x\,\mbox{Tr}\left(-\frac 1 4 F_{\mu \nu}F^{\mu \nu}+\frac i 2
\bar\psi D\!\!\!\!/\psi\right), \label{ac1}
\end{eqnarray}
where $\psi$ is a Majorana-Weyl spinor which transforms in 
the adjoint representation of $U(n)$. The action is invariant under the 
supersymmetry transformations rules
\begin{eqnarray}
\delta A_\mu &=& i  \bar\epsilon\gamma_\mu\psi,\\
\delta\psi&=&-\frac 1 2 F_{\mu \nu}\gamma^{\mu \nu}\epsilon+\eta, \label{dpsi}
\end{eqnarray}
with $A_\mu $ the $U(n)$ gauge potential and $\epsilon$ and $\eta$ constant 
Majorana-Weyl spinors. 

The leading term of the effective theory describing $n$ coinciding 
D$2p$-branes ($p\geq 2$) is nothing but eq. (\ref{ac1}) dimensionally reduced to 
$2p+1$ dimensions. The gauge potentials in the transverse directions appear
as $9-2p$ scalar fields in the adjoint representation of $U(n)$,
which are reinterpreted as the transversal coordinates of the D-branes. 
As they will not play any significant role in this paper, we drop them
from now on.

We now proceed with the analysis of eq. (\ref{dpsi}) in the presence of 
magnetic background fields and demand that some supersymmetry is preserved. 
I.e. we investigate whether for certain magnetic background fields there 
is an $\epsilon $ such that $\delta\psi=0$. In fact we can use the $\eta$ 
transformation in eq. (\ref{dpsi}) to reduce any $F_{\mu \nu}$ from $u(n)$ 
to $su(n)$.
We start by switching on $F_{2i-1\,2i}\in su(n)$, $i\in\{1,\cdots,p\}$, 
which satisfy the BPS condition suggested by D-branes at angles
\begin{eqnarray}
\sum_{i=1}^pF_{2i-1\,2i}=0.\label{bps1}
\end{eqnarray}
For further convenience, 
we switch to complex coordinates (for details, we refer to appendix A),
where we have that $F_{\alpha \bar\alpha }=iF_{2\alpha -1\,2\alpha }$. 
Eq. (\ref{bps1}) becomes in complex coordinates,\footnote{Unless
stated otherwise, we sum over repeated indices.}
\begin{eqnarray}
F_{\alpha \bar\alpha }\equiv\sum_{\alpha } F_{\alpha \bar\alpha } =0.  \label{bps2}
\end{eqnarray}
We get that
\begin{eqnarray}
\delta\psi =F_{\alpha \bar\alpha }\gamma_{\alpha \bar\alpha }\epsilon=0,
\end{eqnarray}
holds provided that
\begin{eqnarray}
\epsilon=\prod_{\alpha =1}^p(1+\gamma_{1\bar 1\alpha \bar\alpha })\xi,
\end{eqnarray}
with $\xi$ an arbitrary Majorana-Weyl spinor. This reduces the number of 
supersymmetry charges from 16 to $16/2^{p-1}$.

It is not hard to check that when {\em all} magnetic fields are switched 
on, $\delta\psi =0$ still holds provided the magnetic fields do not only 
satisfy eq. (\ref{bps2}) but
\begin{eqnarray}
F_{\alpha \beta}=F_{\bar\alpha \bar\beta}=0,\quad \alpha 
,\,\beta\in\{1,\cdots,p\},  \label{bps3}
\end{eqnarray}
as well. For $p=2$, eqs. (\ref{bps2}) and (\ref{bps3}) are nothing 
but the well known instanton equations. In general eq. (\ref{bps3}) 
defines a holomorphic vector bundle while eq. (\ref{bps2}), which can be 
rewritten in a more covariant form,
\begin{eqnarray}
g^{\alpha \bar\beta}F_{\alpha \bar\beta}=0,
\end{eqnarray}
is the Donaldson-Uhlenbeck-Yau condition for stability of the vector
bundle \cite{GSW}.

Remains to check whether these configurations solve the equations of 
motion, $D^\mu F_{\mu \nu}=0$. In complex coordinates, this becomes
\begin{eqnarray}
0&=&D_{\bar\alpha }F_{\alpha \bar\beta}+D_{\alpha }F_{\bar\alpha 
\bar\beta}\nonumber\\
&=&D_{\bar\beta}F_{\alpha \bar\alpha }+2D_{\alpha }F_{\bar\alpha 
\bar\beta}, 
\end{eqnarray}
where we used the Bianchi identities. This is indeed satisfied if 
eqs. (\ref{bps2}) and (\ref{bps3}) hold. Note that magnetic field 
configurations satisfying eqs. (\ref{bps2}) and (\ref{bps3}) always solve 
the equations of motion and always preserve supersymmetry, even when they 
are not constant. As a consequence, we will not demand them to be constant 
anymore.

\setcounter{equation}{0}
\section{Deformations}
A natural question which arises is whether we can deform the Yang-Mills 
action in such a way that the BPS configurations given in the previous 
section remain solutions to the equations of motion. Though the discussion in the
previous section holds for both the abelian as well as the non-abelian case, we focus in
the remainder of this paper on the abelian case. In this way we avoid the 
additional complication of having to take the different ordenings into account. 
{From} now on the magnetic fields
take values in the Cartan subalgebra of $u(n)$ and we postpone the study of 
the non-abelian extension to a future paper \cite{wij}. In addition, we will work 
under the assumption that the fieldstrengths vary slowly. In other 
words, we add terms polynomial in the fieldstrength to the action and 
ignore terms containing derivatives of the fieldstrength (acceleration terms).
We will further comment on these assumptions in the concluding section.
Under these assumptions, we arrive at equations of motion of the form
\begin{eqnarray}
D^\mu F_{\mu \nu}+x\, D^\mu (F_{\mu \rho}F^{\rho\sigma}F_{\sigma\nu})+
y\,D^\mu (F_{\mu \nu}F_{\rho\sigma}F^{\sigma\rho})+{\cal O}(F^5)=0,
\label{eom3}
\end{eqnarray}
where $x$ and $y$ are real constants. As we saw before, the analysis of 
the leading order term led to the conditions (in complex coordinates)
\begin{eqnarray}
F_{\alpha\beta}=F_{\bar\alpha \bar\beta}=0,\label{hol}\\
g^{\alpha \bar\beta}F_{\alpha \bar\beta}= F_{\alpha \bar\alpha 
}=0,\label{duy}
\end{eqnarray}
where in the last line we used the fact that we are working in flat space.
Passing to complex coordinates while implementing the holomorphicity 
conditions eq. (\ref{hol}), eq. (\ref{eom3}) becomes,
\begin{eqnarray}
D_{\bar\alpha}F_{\alpha \bar\beta}+x\,D_{\bar\alpha } (F_{\alpha \bar\gamma} 
F_{\gamma\bar\delta}F_{\delta\bar\beta})+2y\,
D_{\bar\alpha } (F_{\alpha \bar\beta} 
F_{\gamma\bar\delta}F_{\delta\bar\gamma})+{\cal O}(F^5)=0.
\end{eqnarray}
Upon using the Bianchi identities and eq. (\ref{hol}), this results in
\begin{eqnarray}
&&D_{\bar\beta}(F_{\alpha \bar\alpha}+\frac x 3 F_{\alpha 
\bar\gamma}F_{\gamma\bar\delta}F_{\delta\bar\alpha })+(2y+ \frac x 2 ) 
F_{\alpha \bar\beta}D_{\bar\alpha 
}(F_{\gamma\bar\delta}F_{\delta\bar\gamma}) + x 
F_{\gamma\bar\delta}F_{\delta\bar\beta}D_{\bar\gamma}F_{\alpha 
\bar\alpha }  +\nonumber\\&& \qquad 2 y 
F_{\gamma\bar\delta}F_{\delta\bar\gamma}D_{\bar\beta} F_{\alpha \bar\alpha 
}+{\cal O}(F^5)=0,\label{a2}
\end{eqnarray}
which vanishes if
\begin{eqnarray}
y=-\frac x 4 ,
\end{eqnarray}
holds and provided that we deform the Donaldson-Uhlenbeck-Yau condition,
eq. (\ref{duy}), to
\begin{eqnarray}
F_{\alpha \bar\alpha}+\frac x 3 F_{\alpha 
\bar\gamma}F_{\gamma\bar\delta}F_{\delta\bar\alpha } +{\cal O}(F^5)=0.
\label{duy1}
\end{eqnarray}
Rescaling $F$ and multiplying the equation of motion with a constant,
we can put $x=1$. Upon restoring the $SO(2p)$ 
invariance, we find that the equations of motion integrate to the action
\begin{eqnarray}
{\cal S}&=&\int d^{2p+1}x\left(\frac 1 4 F_{\mu_1 \mu_2}F^{\mu_2 \mu_1}+ 
\frac 1 8 F_{\mu_1 \mu_2}F^{\mu_2\mu_3}F_{\mu _3\mu _4}F^{\mu _4\mu 
_1}\right.
\nonumber\\
&&\left. -
\frac{1}{32}(F_{\mu_1 \mu_2}F^{\mu_2 \mu_1})^2+{\cal O}( F^6) \right), 
\end{eqnarray}
which, modulo an undetermined overall multiplicative constant,
we recognize as the Born-Infeld action through order $F^4$ (see appendix B).

In a similar way, one can push this calculation an order higher by adding
the most general integrable terms through fifth order in $F$ to the
equations of motion. Again we require that the (deformed) BPS solutions
solve the equations of motion. The scale of the fieldstrengths was
already fixed at previous order. In this calculation one needs e.g. that
the two last terms in eq. (\ref{a2}) get completed to a derivative of eq.
(\ref{duy1}). At the end one finds that the equations of motion get
uniquely fixed and they indeed integrate to the Born-Infeld action
through sixth order in $F$. Furthermore the Donaldson-Uhlenbeck-Yau
condition acquiers an order $F^5$ correction,
\begin{eqnarray}
F_{\alpha \bar\alpha}+\frac 1 3 F_{\alpha 
\bar\gamma}F_{\gamma\bar\delta}F_{\delta\bar\alpha } +
\frac 1 5 F_{\alpha 
\bar\gamma}F_{\gamma\bar\delta}F_{\delta\bar\epsilon}
F_{\epsilon\bar\zeta}F_{\zeta\bar\alpha }
+{\cal O}(F^7)=0.
\end{eqnarray}

These results raise the suspicion that the Born-Infeld action is the {\em 
only} deformation of Yang-Mills which allows for BPS solutions of the form 
eqs. (\ref{hol}-\ref{duy}). Furthermore one expects that
the holomorphicity conditions, eq. (\ref{hol}) remain unchanged, 
while the Donaldson-Uhlenbeck-Yau condition, eq. (\ref{duy}) receives 
$\alpha '$ corrections. In the next section, we will show that this is 
indeed the case.

\setcounter{equation}{0}
\section{All order results}

In this section we will construct the unique deformation of abelian Yang-Mills
which allows for BPS solutions which are in leading order given by eqs. 
(\ref{hol}-\ref{duy}).
Consider a general term in the deformed Yang-Mills lagrangian,
\begin{eqnarray}
\lambda_{(p_1,p_2,\ldots,p_n)} \, (\tr F^2)^{p_1} (\tr F^4)^{p_2} 
\ldots (\tr F^{2n})^{p_n} \, ,\quad p_i\in\IN,\ \forall i 
\in\{1,\cdots,n\},
\end{eqnarray}
where $\lambda_{(p_1,p_2,\ldots,p_n)}\in\IR$.
Dropping the overall $\lambda_{(p_1,\ldots,p_n)}$, this term contributes 
to the equations of motion by,
\begin{eqnarray}
\label{gentermeom}
\sum_{j=1}^{n} 4 j \, p_j D^{\mu} \left( \left(F^{2j-1}\right)_{\mu\nu}
        (\tr F^2)^{p_1} (\tr F^4)^{p_2} \ldots (\tr F^{2j})^{p_j - 1} 
        \ldots (\tr F^{2n})^{p_n} \right) \, .
\end{eqnarray}
Passing to complex coordinates, eq. (\ref{cc}), we get
\begin{eqnarray}
\label{gentermcomp}
\sum_{j=1}^{n} 4 j \, p_j \, 2^{p_1+\cdots+p_n-1} D_{\bar{\alpha}} 
\left( \left(F^{2j-1}\right)_{\alpha\bar{\beta}}
      (F^2)^{p_1} \ldots (F^{2j})^{p_j-1} \ldots (F^{2n})^{p_n} \right) \, ,
\end{eqnarray}
where
\begin{eqnarray} 
(F^{m})_{\alpha\bar{\beta}} & \equiv& F_{\alpha\bar{\alpha}_2} F_{\alpha_2\bar{\alpha}_3} 
\ldots F_{\alpha_m\bar{\beta}} \nonumber\\
(F^{m}) & \equiv&  F_{\alpha_1\bar{\alpha}_2} F_{\alpha_2\bar{\alpha}_3} 
\ldots F_{\alpha_m\bar{\alpha}_1} . 
\end{eqnarray}
Using the Bianchi identies and eq. (\ref{hol}), we find for the action of the 
derivative operator $D_{\bar{\alpha}}$ on $(F^{2j-1})_{\alpha\bar{\beta}}$:
\begin{eqnarray}
D_{\bar{\alpha}} (F^{2j-1})_{\alpha\bar{\beta}}
 = \sum_{h=1}^{2j-2} \frac{1}{h} \left( D_{\bar{\alpha}} 
F^{h} \right) (F^{2j-1-h})_{\alpha\bar{\beta}}+
\frac{1}{2j-1}D_{\bar\beta}(F^{2j-1})\, .
\end{eqnarray}
Implementing this result in eq. (\ref{gentermcomp}) yields,
\begin{eqnarray}
\label{gentermexp}
&&\sum_{j=1}^{n} 4 j \, p_j \, 2^{p_1+\cdots+p_n-1}
 \Biggl( \biggr(\sum_{h=1}^{2j-2} \frac{1}{h}
\left( D_{\bar{\alpha}} F^{h} \right)(F^{2j-1-h})_{\alpha\bar{\beta}} 
 + \frac{1}{2j-1}  
 D_{\bar\beta}(F^{2j-1})\biggr) \nonumber\\
 &&
(F^2)^{p_1}  \ldots (F^{2j})^{p_j-1} \ldots (F^{2n})^{p_n} + 
\nonumber\\
&& \sum_{g=1, \; g \neq j}^n p_g \left(D_{\bar{\alpha}} F^{2g}\right)
\left(F^{2j-1}\right)_{\alpha\bar{\beta}} (F^2)^{p_1} \ldots (F^{2j})^{p_j-1} \ldots
(F^{2g})^{p_g-1}\ldots (F^{2n})^{p_n}+  \nonumber\\
&& (p_j -1) \left(D_{\bar{\alpha}} F^{2j}\right)
\left(F^{2j-1}\right)_{\alpha\bar{\beta}} (F^2)^{p_1} \ldots (F^{2g})^{p_j-2} 
\ldots (F^{2n})^{p_n}  \Biggr)\, .
\end{eqnarray}
We now study the different types of terms in the equations of motion. We will determine
the numerical prefactors such that the (deformed) BPS configurations solve them.

\begin{itemize}
\item Terms of the form $D_{\bar{\beta}} F^{2r-1}$: there is one of 
these terms in each order and they add up to,
\begin{eqnarray}
D_{\bar{\beta}} \left( 4 \cdot 1 \lambda_{(1)} F_{\alpha \bar\alpha } + 
\frac{4 \cdot 2}{3}\lambda_{(0,1)} (F^3)_{\alpha \bar\alpha } + 
\frac{4 \cdot 3}{5} \lambda_{(0,0,1)} 
(F^5 )_{\alpha \bar\alpha }
+ \cdots \right) \, .
\end{eqnarray}
In leading order it vanishes because of eq. (\ref{duy}). 
It is clear that the all order expression should vanish by itself 
thereby giving the deformed Donaldson-Uhlenbeck-Yau condition,
\begin{eqnarray}
\frac 1 1 \lambda_{(1)} F_{\alpha \bar\alpha } + 
\frac{ 2}{3}\lambda_{(0,1)} (F^3)_{\alpha \bar\alpha } + 
\frac{ 3}{5} \lambda_{(0,0,1)} 
(F^5 )_{\alpha \bar\alpha }
+ \cdots =0\, .  \label{duyg}
\end{eqnarray}
\item Terms of the form $\left( D_{\bar{\alpha}} F^{2r} \right) 
(F^{2l-1})_{\alpha\bar{\beta}} \tail$, where
\begin{eqnarray}
\tail= (F^2)^{p_1} (F^4)^{p_2}\ldots (F^{2n})^{p_n}=\prod_{i=1}^n(F^{2i})^{p_i}:
\end{eqnarray} 
as these terms involve traces over even powers of the fieldstrength they 
can never be cancelled by a condition as eq. (\ref{duyg}), so they should 
cancel order by order among themselves.
If we look at the first versus the two last terms of 
eq. (\ref{gentermexp}), we see immediately that a term of this form originates
from two different terms in the action, namely $(\tr F^{2l+2r}) \tail$ and
$(\tr F^{2l}) (\tr F^{2r})\tail$.  Suppose first that $l \neq r$.
Requiring such a term to vanish results, using the first two terms
in eq. (\ref{gentermexp}), in the following condition,
\begin{eqnarray}
\label{coeffcond1}
&&(l+r)(p_{l+r}+1) \lambda_{(\ldots,p_l,\ldots,p_r,\ldots,p_{l+r}+1,
\ldots)} +\nonumber\\
&&4 l r(p_l+1)(p_r+1) \lambda_{(\ldots,p_l+1,\ldots,p_r+1,
\ldots,p_{l+r},\ldots)} =0 \, .
\end{eqnarray}
The Born-Infeld coefficients eq. (\ref{BIfactors}) satisfy this condition.
Analogously, using the first and third term in eq. (\ref{gentermexp}), 
we get when $l = r$,
\begin{eqnarray}
\label{coeffcond2}
(p_{2l}+1) \lambda_{(\ldots,p_l,\ldots,p_{2l}+1,\ldots)}+
2 l (p_l+2)(p_l+1) \lambda_{(\ldots,p_l+2,\ldots,p_{2l},\ldots)} =0 \, ,
\end{eqnarray}
again satisfied by the Born-Infeld coefficients eq. (\ref{BIfactors}).
Note that the two conditions eq. (\ref{coeffcond1}) and eq. (\ref{coeffcond2}) 
are enough to determine all coefficients at a
certain order if one is known.  
We give an example of the chain of relations at order $F^{8}$,

\setlength{\unitlength}{1mm}

\begin{center}
\begin{picture}(120,25)
\put(10,20){\makebox(0,0){$(4,0,0,0)$}}
\put(20,20){\vector(1,0){10}}
\put(25,22){\makebox(0,0)[b]{$\scriptstyle{(\ref{coeffcond2})}$}}
\put(40,20){\makebox(0,0){$(2,1,0,0)$}}
\put(50,20){\vector(1,0){10}}
\put(55,22){\makebox(0,0)[b]{$\scriptstyle{(\ref{coeffcond2})}$}}
\put(70,20){\makebox(0,0){$(0,2,0,0)$}}
\put(80,20){\vector(1,0){10}}
\put(85,22){\makebox(0,0)[b]{$\scriptstyle{(\ref{coeffcond2})}$}}
\put(100,20){\makebox(0,0){$(0,0,0,1)$}}
\put(40,2){\makebox(0,0){$(1,0,1,0)$}}
\put(40,6){\vector(0,1){10}}
\put(42,11){\makebox(0,0)[l]{$\scriptstyle{(\ref{coeffcond1})}$}}
\put(50,2){\vector(3,1){42}}
\put(71,11){\makebox(0,0)[r]{$\scriptstyle{(\ref{coeffcond1})}$}}
\end{picture}
\end{center}

So, up until now, we find Born-Infeld modulo a proportionality factor at each order,
\begin{eqnarray}
\label{factorsx}
\lambda_{(p_1,p_2,\ldots,p_n)}=\frac{(-1)^{k+1}}{4^k} \frac{1}{p_1 ! \ldots p_n !} \,
                               \frac{1}{1^{p_1} \ldots n^{p_n}} \, X_{\sum j p_j},
\end{eqnarray}
where $X_{\sum_{j=1}^n j p_j}\in\IR$ are unknown constants.
\item Terms of the form $\left( D_{\bar{\beta}} F^{2r-1} \right)  \tail$: 
they relate different 
orders in $F$. The only way to cancel these terms is by virtue of eq. (\ref{duyg}). 
Using eqs. (\ref{gentermexp}) and (\ref{factorsx}) we find that such a term appears 
in the equation of motion as
\begin{eqnarray}
\frac{(-1)^{\sum_l p_l}}{2^{\sum_l p_l}
\prod_l (p_l!)\prod_l l^{p_l}}\frac{X_{\sum_l l p_l+r}}{2r-1}
\left( D_{\bar{\beta}} F^{2r-1} \right)  \tail ,
\end{eqnarray}
where all summations and products run from $l=1$ through $l=n$. For a 
given tail, the sum over $r$ of such terms has to vanish through the use
of eq. (\ref{duyg}). This determines all unknowns $X_r$ in terms of two,
\begin{eqnarray}
X_r=X_2\left(\frac{X_2}{X_1}\right)^{r-2},\quad r\geq 3.\label{rels}
\end{eqnarray}
We still have the freedom to rescale the fieldstrength in the equations of 
motion by an arbitrary factor and we also note that the equations of 
motion are only determined modulo an arbitrary multiplicative factor. In 
other words, we can only determine the action modulo an overall 
multiplicative factor. This freedom can be used to put $X_1=X_2=1$. 
Combining this with eq. (\ref{rels}), we get 
\begin{eqnarray}
X_r=1, \quad\forall r \geq 1. \label{ctfix}
\end{eqnarray}
At this point the equations of motion are completely fixed and they are 
exactly equal to the equations of motion of the abelian Born-Infeld 
theory, implying that our action is, modulo an overall multiplicative 
constant, the Born-Infeld action. This fixes the BPS condition, eq. 
(\ref{duyg}), as well,
\begin{eqnarray}
\label{pertinst}
0 & = & F_{\alpha \bar\alpha } + \frac{1}{3} (F^3)_{\alpha \bar\alpha } + 
\frac{1}{5} (F^5)_{\alpha \bar\alpha } + \cdots \nonumber\\
  & =&\mbox{tr}\, {\rm arctanh} \, {\cal F},
\end{eqnarray}
where ${\cal F}$ is a $p\times p$ matrix with elements $F_{\alpha 
\bar\beta}$ and the trace is taken over the Lorentz indices.
\item Terms of the form $\left( D_{\bar{\alpha}} F^{2r-1} \right) 
(F^{2s})_{\alpha\bar{\beta}} \tail$: 
these are the only terms left and they will cancel because of eq. 
(\ref{pertinst}). Using
eqs. (\ref{gentermexp}), (\ref{factorsx}) and (\ref{ctfix}), 
we get the prefactor of such term, 
\begin{eqnarray}
\frac{(-1)^{\sum_l p_l}}{2^{\sum_l p_l}
\prod_l (p_l!)\prod_l l^{p_l}}\frac{1}{2r-1}
\left( D_{\bar{\alpha}} F^{2r-1} \right) (F^{2s})_{\alpha\bar{\beta}} \tail,
\end{eqnarray}
and it is clear that when summing over $r$ they vanish because of eq. 
(\ref{pertinst}).
\end{itemize}
This completes the proof that the abelian Born-Infeld action is the
unique deformation of the abelian Yang-Mills action which allows for
BPS solutions.

\setcounter{equation}{0}
\section{Discussion and conclusions}
Fieldstrength configurations which define a stable, eq. (\ref{duy}),
holomorphic, eq. (\ref{hol}), vector bundle solve the Yang-Mills
equations of motion. Such configurations are relevant in the study of BPS
solutions for D-branes in the $\alpha '\rightarrow 0$ limit. In this
paper we deformed the {\it abelian} theory by adding arbitrary powers of
the fieldstrength to the Yang-Mills lagrangian. Demanding that a
deformation of eqs. (\ref{hol}-\ref{duy}) still solves the equations of
motion, we showed that the deformation is uniquely determined: it is
precisely the abelian Born-Infeld theory. The holomorphicity condition
eq. (\ref{hol}) remains unchanged while the Donaldson-Uhlenbeck-Yau
stability condition gets deformed to 
\begin{eqnarray} \mbox{tr}\, {\rm arctanh} \, {\cal F}=0,\label{duyff} 
\end{eqnarray} 
with ${\cal F}$ a $p\times p$ matrix with elements $F_{\alpha
\bar\beta}$. 

The analysis in section 5 holds not only in flat space but in K\"ahler
geometries as well. Defining ${\cal F}$ to be the $p\times p$ matrix with
elements $F_\alpha {}^\beta \equiv g^{\beta\bar\gamma}F_{\alpha\bar\gamma
}$ with $g_{\alpha \bar\beta}$ the K\"ahler metric, one finds again eq.
(\ref{duyff}). In this context, it might be worthwhile to mention that it
would be interesting to include the transverse scalars in the analysis.
This would make it possible to get an all $\alpha '$ result for the
stability condition for branes wrapped around a holomorphic submanifold
of a K\"ahler manifold \cite{HM}.

Eqs. (\ref{hol}) and (\ref{duyff}) play an important role in the study of
BPS configurations for D-branes at finite $\alpha '$. As supersymmetry
and magnetic field configurations discussed above are closely related,
our result provides evidence strengthening the belief that the only
supersymmetric deformation of ten-dimensional supersymmetric $U(1)$
Yang-Mills theory is the supersymmetric Born-Infeld action.

Eq. (\ref{duyff}) holds in any dimension $d=2p$. It is a $U(p)$ invariant 
and therefore it can be rewritten in terms of Casimir invariants. As 
$U(p)$ has Casimir invariants of order 1, 2, ..., $p$, one can rewrite eq. 
(\ref{duyff}) in a less elegant though more familiar form 
when specifying to particular
dimensions. We tabulate the resulting equivalent expressions for the
cases relevant to D-brane physics.

\begin{center}
\begin{tabular}{|c|c|}\hline\hline
$p$ & stability condition\\ \hline\hline
2 &$F_{1\bar 1}+F_{2\bar 2}=0$ \\ \hline
3 &$F_{1\bar 1}+F_{2\bar 2}+F_{3\bar 3}+
F_{1\bar 1}F_{2\bar 2}F_{3\bar 3}
-F_{1\bar 2}F_{2\bar 1}F_{3\bar 3}
-F_{1\bar 3}F_{2\bar 2}F_{3\bar 1}
+ F_{1\bar 2}F_{2\bar 3}F_{3\bar 1}$\\
&$+F_{1\bar 3}F_{2\bar 1}F_{3\bar 2}
-F_{1\bar 1}F_{2\bar 3}F_{3\bar 2} =0 $\\ \hline
4 &$F_{1\bar 1}+F_{2\bar 2}+F_{3\bar 3}+F_{4\bar 4}
-F_{1\bar 3} F_{2\bar 2} F_{3\bar 1}
+F_{1\bar 2} F_{2\bar 3} F_{3\bar 1}
+F_{1\bar 4} F_{4\bar 3} F_{3\bar 1 }
-F_{1\bar 3} F_{4\bar 4} F_{3\bar 1} $\\
&$+F_{1\bar 3} F_{2\bar 1} F_{3\bar 2 }
-F_{1\bar 1 }F_{2\bar 3} F_{3\bar 2 }
- F_{1\bar 2} F_{2\bar 1} F_{3\bar 3}
+F_{1\bar 1} F_{2\bar 2} F_{3\bar 3 }
-F_{1\bar 4} F_{2\bar 2} F_{4\bar 1 }
+F_{1\bar 2} F_{2\bar 4 }F_{4\bar 1}$\\
&$-F_{1\bar 4} F_{3\bar 3} F_{4\bar 1}
+F_{1\bar 3} F_{3\bar 4} F_{4\bar 1}
+ F_{1\bar 4}F_{2\bar 1} F_{4\bar 2}
-F_{1\bar 1} F_{2\bar 4} F_{4\bar 2}
-F_{2\bar 4} F_{3\bar 3} F_{4\bar 2 }
+F_{2\bar 3} F_{3\bar 4} F_{4\bar 2 }$\\
&$+F_{2\bar 4} F_{3\bar 2} F_{4\bar 3}
-F_{1\bar 1}F_{3\bar 4} F_{4\bar 3}
-F_{2\bar 2} F_{3\bar 4} F_{4\bar 3}
-F_{1\bar 2} F_{2\bar 1} F_{4\bar 4}
+F_{1\bar 1} F_{2\bar 2} F_{4\bar 4}
-F_{2\bar 3} F_{3\bar 2} F_{4\bar 4}$\\
&$+F_{1\bar 1 }F_{3\bar 3} F_{4\bar 4}
+F_{2\bar 2} F_{3\bar 3} F_{4\bar 4}
=0$ 
\\ \hline\hline
\end{tabular}
\end{center}

\setlength{\unitlength}{1mm}

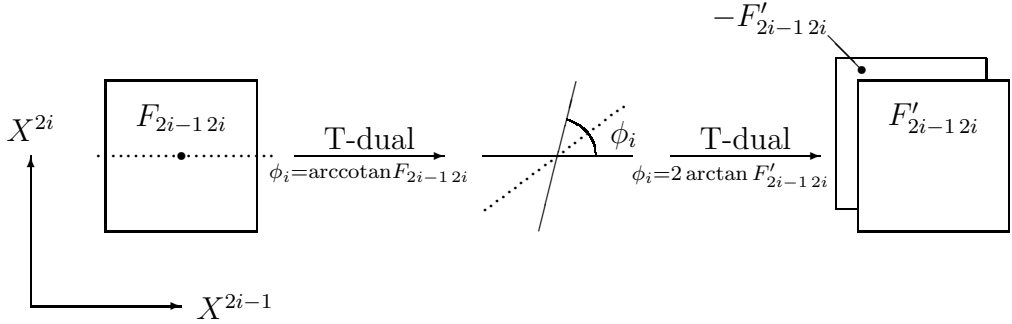
\begin{figure}[tbp]
\centering
\setlength{\unitlength}{1mm}
\begin{picture}(135,50)(5,5)
\put(5,5){\vector(0,1){20}}
\put(5,5){\vector(1,0){20}}
\put(27,5){\makebox(0,0)[l]{$X^{2i-1}$}}
\put(5,27){\makebox(0,0)[b]{$X^{2i}$}}
\put(15,15){\framebox(20,20){}}
\put(25,30){\makebox(0,0){$F_{2i-1\,2i}$}}
\put(25,25){\circle*{1}}
\multiput(14,25)(1,0){24}{\circle*{0.2}}
\put(40,25){\vector(1,0){20}}
\put(50,26){\makebox(0,0)[b]{T-dual}}
\put(50,24){\makebox(0,0)[t]{$\scriptstyle{\phi_i={\rm arccotan} F_{2i-1\,2i}}$}}
\put(65,25){\line(1,0){20}}
\put(72.5,15){\line(1,4){5}}
\multiput(65.4,17.8)(0.8,0.6){24}{\circle*{0.2}}
\qbezier(80,25)(80,28.5)(76.21,29.85)
\put(82,26){\makebox(0,0)[bl]{$\phi_i$}}
\put(90,25){\vector(1,0){20}}
\put(100,26){\makebox(0,0)[b]{T-dual}}
\put(98,24){\makebox(0,0)[t]{$\scriptstyle{\phi_i=2\arctan F'_{2i-1\,2i}}$}}
\put(115,15){\framebox(20,20){}}
\put(125,30){\makebox(0,0){$F'_{2i-1\,2i}$}}
\put(111,41){\makebox(0,0)[br]{$-F'_{2i-1\,2i}$}}
\put(111,41){\line(+1,-1){4.5}}
\put(115.5,36.5){\circle*{1}}
\put(112,18){\line(1,0){3}}
\put(112,18){\line(0,1){20}}
\put(112,38){\line(1,0){20}}
\put(132,35){\line(0,1){3}}
\end{picture}
\vspace{.2cm}
\caption{{\small Before T-dualizing, we have a D-brane extending in the $2i-1\,2i$
plane with $u(1)$ magnetic flux $F_{2i-1\,2i}$ and another D-brane without magnetic flux
perpendicular to it. The dotted lines show the directions along which we T-dualize. 
After T-dualizing twice, we end up with two D-branes which coincide in the $2i-1\,2i$
plane with an $su(2)$ flux $F'_{2i-1\,2i}\sigma_3$. This shows that the
more exotic BPS conditions minimalizing the energy
are T-dual to the ones studied in this paper.}}
\label{braneduality}
\end{figure}

We expect that these BPS solutions minimize the energy. Let us briefly
investigate this for the case where eq. (\ref{hol})
holds\footnote{Throughout this discussion, we assume that the rhs of the
inequalities are differences of invariants. This is certainly so for
$p=4$, see e.g. \cite{GSW}. For $p\geq 6$ this is very probably true as
well. We postpone a more detailed examination of the energy of BPS
configurations to a future publication.}. 
For simplicity, we will only 
switch on the magnetic fields $F_{1\bar 1}$, $F_{2\bar 2}$, ..., $F_{p\bar 
p}$. The energy is given by
\begin{eqnarray}
E=\int d^4 x \big|\det \left({\bf 1}-{\cal F}\right)\big|  =
\int d^4x \prod_{\alpha =1}^p 
\big| \left(1-F_{\alpha \bar\alpha }\right)\big|.
\end{eqnarray}
For $p=2$ we get,
\begin{eqnarray}
E=\int d^4x\big|\left(1-F_{1\bar 1}- F_{2\bar 2}+F_{1\bar 1}F_{2\bar 2}\right)\big| \geq 
\int d^4 x \big| \left|1+F_{1\bar 1}F_{2\bar 2}\right|-\left|F_{1\bar 1}+ 
F_{2\bar 2} \right|\big|.
\end{eqnarray}
Contrary to the Yang-Mills case, we find two situations in which the relation
gets saturated. The first is when $F_{1\bar 1}+ F_{2\bar 2}=0$, which we 
recognize as the familiar BPS condition we have been discussing so far. 
The second configuration is characterized by $1+F_{1\bar 1}F_{2\bar 2} 
=0$. This corresponds to a D2/D4 system. Though this system is not 
supersymmetric, it becomes so when we switch on magnetic fields $F_{1\bar 1}$ and
$F_{2\bar 2}$ on the 
D4-brane which precisely satisfy $1+F_{1\bar 1}F_{2\bar 2} =0$.
For $p=3$, we get 
\begin{eqnarray}
E&=&\int d^6x\big| \left(1-F_{1\bar 1}- F_{2\bar 2} - F_{3\bar 3} 
+F_{1\bar 1}F_{2\bar 2} +F_{1\bar 1}F_{3\bar 3}+F_{2\bar 2}F_{3\bar 3}
- F_{1\bar 1}F_{2\bar 2}F_{3\bar 3} 
\right)\big| \nonumber\\
&\geq& 
\int d^6x\big| \left|1+F_{1\bar 1}F_{2\bar 2}+F_{1\bar 1}F_{3\bar 3}+F_{2\bar 2}F_{3\bar 3}
\right|-\nonumber\\
&&\qquad\left|F_{1\bar 1}+ F_{2\bar 2} + F_{3\bar 3}+
F_{1\bar 1}F_{2\bar 2}F_{3\bar 3}
\right|\big|.  \label{p3}
\end{eqnarray}  
Again the result is saturated in two cases. When the last factor vanishes 
in eq. (\ref{p3}) we have the standard BPS condition. When the first factor 
vanishes, we find a configuration corresponding to either a D0/D6 or a 
D4/D6 system. By switching on magnetic fields $F_{1\bar 1}$, $F_{2\bar 2}$ 
and $F_{3\bar 3}$ on the D6-brane which satisfy this relation we obtain a 
BPS configuration. 
A similar analysis holds for $p=4$. Either one recovers the standard BPS 
configuration of D8-branes or a D2/D8 system (or equivalently a D6/D8 
system) with magnetic fields on the D8-brane such that the result is BPS.
Aspects of some of these non-standard BPS configurations were studied in
\cite{WittenBPS} .
Even as these ``exotic'' BPS configurations have no $\alpha '\rightarrow 0$
limit, they are in fact T-dual to the BPS configurations studied in section 
2 as is demonstrated in figure \ref{braneduality}. 

Our analysis was performed under the assumption that the fieldstrengths
vary slowly, i.e. we ignored terms having derivatives of the
fieldstrength. Such terms are expected to be present \cite{AT}. It would
be very interesting to investigate whether our method can handle such
terms as well. However, an additional complication will arise in such an
analysis. As explained in \cite{AT2} and \cite{GW}, because of field
redefinitions, derivative terms are ambiguous. Nonetheless, it is
worthwile to investigate this point as this will further clarify the
relation between the commutative and non-commutative pictures \cite{SW},
\cite{CS}.

Another point which deserves further attention is the study of the BPS
conditions as a function of the string coupling constant $g_S$. In this
way the method developed in this paper might provide an alternative
approach to the study of the effective action as a function of the string
coupling constant. In \cite{AT2}, it was shown that through second order
in $g_S$ and in flat space the Born-Infeld action, modulo a
renormalization of the tension, still describes the effective dynamics.
It would be intriguing if such a claim could be pushed at higher orders
(note that in non-trivial geometries this is very probably not true).

Finally, the results in this paper provide sufficient motivation for a
detailed investigation of the non-abelian case. As eqs. (\ref{hol}) and
(\ref{duy}) hold both in the abelian and the non-abelian case, we can
still use it as a starting point and investigate allowed deformations.
Not only do we expect a concrete ordening prescription for the action,
but eq. (\ref{duyff}) should get supplied with an ordening prescription
as well. Note that derivative terms might become relevant in this case.
We will report on this in \cite{wij}.

\vspace{5mm}

\noindent {\bf Acknowledgments}:
We thank Eric Bergshoeff, Frederik Denef, Mees de Roo, Marc Henneaux,
Walter Troost and Michel Van den Bergh for useful discussions. In
particular, we are grateful to Jan Troost for numerous suggestions and
illuminating conversations. This work is supported in part by the
FWO-Vlaanderen and in part by the European Commission RTN programme
HPRN-CT-2000-00131, in which the authors are associated to the university
of Leuven.

\vspace{5mm}

\appendix

\setcounter{equation}{0}
\section{Notations and conventions}
Our metric follows the ``mostly plus'' conventions. Indices denoted
by $\mu $, $\nu$, ... run from 0 
to $2p$, denoted by $i$, $j$, ... run from 1 to $2p$ and denoted by
$\alpha $, $\beta$, ... run from 1 to $p$. We use real $32\times32$ 
$\gamma$-matrices satisfying $\{\gamma_\mu ,\gamma_\nu\}=2g_{\mu \nu}$ and 
$\gamma_\mu ^T=\gamma_0\gamma_\mu \gamma_0$. By $\gamma_{\mu_1\cdots\mu 
_n}$ we denote the (weighted) completely antisymmetrized product 
$\gamma_{[\mu _1}\gamma_{\mu _2} \cdots\gamma_{\mu _n]}$ with $[[\cdots ]] 
= [\cdots ]$.

Instead of using real spatial coordinates $x^i$, $i\in\{1,\cdots,2p\}$, we 
will often use complex coordinates $z^\alpha $, $\alpha \in\{1,\cdots 
p\}$,
\begin{eqnarray}
z^\alpha \equiv \frac{1}{\sqrt 2}\left(x^{2\alpha -1}+ix^{2\alpha }\right),\quad
\bar z^{\bar\alpha} \equiv \frac{1}{\sqrt 2}
\left(x^{2\alpha -1}-ix^{2\alpha }\right).
\label{cc}
\end{eqnarray}
As we work in flat space, the metric is $g_{\alpha \beta}=g_{\bar\alpha 
\bar\beta}=0$, $g_{\alpha \bar\beta}=\delta_{\alpha \bar\beta}$.

Consider the rotation group $SO(p)$. The subgroup preserving the complex 
structure is $U(p)$. If we denote the $so(2p)$ generators by 
$M_{ij}=-M_{ji}$, the $u(p)$ generators are given by the subset $M_{\alpha 
\bar\beta}$. The $u(1)$ generator commuting with all the $u(p)$ generators 
is given by $\sum_{\alpha }M_{\alpha \bar\alpha }$. The remainder of the 
$so(2p)$ generators, $M_{\alpha \beta}$ and $M_{\bar\alpha \bar\beta}$
resp.,
transforms in the $p(p-1)/2$ and the $\overline{p(p-1)/2}$ of $su(p)$ 
resp.

\setcounter{equation}{0}
\section{The abelian Born-Infeld action}
In this appendix we derive a few properties of the abelian Born-Infeld 
action needed in section five.

The Born-Infeld lagrangian can be rewritten as\footnote{The trace denotes 
a trace over the Lorentz indices.}
\begin{eqnarray}
{\cal L}_{BI}&=&-\sqrt{\det(\delta^\mu {}_\nu-F^\mu {}_\nu)}\nonumber\\
&=&\sum_{k=0}^{\infty} \frac{(-1)^{k+1}}{4^k k!} (\tr F^2 + 
\frac{1}{2}\tr F^4 + \cdots +
           \frac{1}{p} \tr F^{2p} + \cdots )^k \, .
\end{eqnarray}
A general term in the abelian Born-Infeld Lagrangian
\begin{eqnarray}
\lambda_{(p_1,p_2,\ldots,p_n)}^{BI} \, (\tr F^2)^{p_1} (\tr F^4)^{p_2} 
\ldots (\tr F^{2n})^{p_n} \, ,
\end{eqnarray}
originates from the $k$th term in the Taylor expansion, 
with $k$ given by
\begin{eqnarray}
k=p_1 + p_2 + \cdots + p_n \, .
\end{eqnarray}
Hence, the numerical prefactor becomes
\begin{eqnarray}
\label{BIfactors}
\lambda_{(p_1,p_2,\ldots,p_n)}^{BI}=\frac{(-1)^{k+1}}{4^k} 
\frac{1}{p_1 ! \ldots p_n !} \, \frac{1}{1^{p_1} \ldots n^{p_n}} \, .
\end{eqnarray}

\end{document}